\documentclass[%
 reprint,
superscriptaddress,
showpacs,preprintnumbers,
nofootinbib,
 amsmath,amssymb,
 aps,
 prc,
floatfix,
]{revtex4-1}

\usepackage{graphicx}
\usepackage{dcolumn}
\usepackage{bm}

\begin{document}
\title{Multipole modes of excitation in triaxially deformed superfluid nuclei}

\author{Kouhei Washiyama}
\email{washiyama@nucl.ph.tsukuba.ac.jp} 
\affiliation{Center for Computational Sciences, 
University of Tsukuba, Tsukuba 305-8577, Japan}

\author{Takashi Nakatsukasa}
\affiliation{Center for Computational Sciences, 
University of Tsukuba, Tsukuba 305-8577, Japan}
\affiliation{Faculty of Pure and Applied Sciences,
University of Tsukuba, Tsukuba 305-8571, Japan}
\affiliation{iTHES, RIKEN, Wako 351-0198, Japan}

\date{\today}

\begin{abstract}
\begin{description}
\item[Background] 
The five-dimensional quadrupole collective model
based on energy density functionals (EDF) has often been employed
to treat long-range correlations associated with shape fluctuations in nuclei. 
Our goal is to derive the collective inertial functions 
in the collective Hamiltonian
by the local quasiparticle random phase approximation (QRPA)
that correctly takes into account time-odd mean-field effects.
Currently, practical framework to perform the
QRPA calculation with the modern EDFs on the $(\beta,\gamma)$ deformation space
is not available.
\item[Purpose] Toward this goal, we develop an efficient numerical method
to perform the QRPA calculation on the $(\beta,\gamma)$ deformation space
based on the Skyrme EDF.
\item[Methods] We use the finite amplitude method (FAM) for efficient
calculation of QRPA strength functions for multipole external fields. 
We construct a computational code of FAM-QRPA in the three-dimensional 
Cartesian coordinate space to handle triaxially deformed superfluid nuclei.
\item[Results] We validate our new code by comparing our results
with former QRPA calculations for axially symmetric nuclei.
Isoscalar quadrupole strength functions in triaxial superfluid nuclei,
${}^{110}$Ru and ${}^{190}$Pt, are obtained
within a reasonable computational cost.
\item[Conclusions] 
QRPA calculations for triaxially deformed superfluid nuclei 
based on the Skyrme EDF
are achieved with the help of FAM.
This is an important step toward the microscopic calculation of
collective inertial functions of the local QRPA.
\end{description}
\end{abstract}
\maketitle

\textit{Introduction.}
%
The shape of atomic nuclei is strongly influenced
by quantum nature of nuclear systems.
Excitation spectra and their transition probabilities
clearly indicate the existence of 
shape fluctuations and shape coexistence \cite{heyde11},
particularly in transitional regions from spherical to deformed
shapes in the ground state.
The long-lived fission products (LLFP) from uranium fueled reactors,
such as Pd and Zr isotopes, are located in transitional regions on
the nuclear chart and demonstrate the shape mixing and coexistence.
It is important to understand basic properties of the LLFPs to
develop a possible nuclear transmutation method,
which is a main target of an ImPACT program
``Reduction and Resource Recycling of High-level Radioactive Wastes
through Nuclear Transmutation'' \cite{impact}.

One of the standard methods of investigating nuclear many-body problems
is the nuclear energy density functional (EDF) theory \cite{bender03}.
The nuclear EDF well describes the ground-state properties of atomic nuclei.
However, in the mean-field level,
it cannot describe shape fluctuations and shape coexistence.
We need to go beyond mean field for description of such phenomena,
including quantum fluctuations associated with
the large-amplitude collective motion.
If the EDF were constructed as expectation value of a well-defined Hamiltonian,
a possible extension would be
the generator coordinate method (GCM) \cite{bender08,rodriguez10,yao10}.
However, most of EDFs are known to have a singular behavior 
\cite{anguiano01,dobaczewski07},
which prevents us from the straightforward application of the GCM.

A practical alternative to the GCM may be
the collective Hamiltonian method.
The five-dimensional quadrupole collective Hamiltonian,
with quadrupole deformation parameters $(\beta, \gamma)$
and three Euler angles,
is constructed from the EDF-based calculation of Skyrme, Gogny,
and covariant EDFs \cite{prochniak04, niksic09, delaroche10}.
The collective potential is obtained by the constrained minimization of
the EDF,
while for the collective inertial functions,
the Inglis--Belyaev cranking formula is employed.
Therefore, the time-odd components in the mean field 
are neglected in those studies,
and an empirical enhancement factor of $1.2-1.4$ is often adopted
for the collective inertias.

Starting from the adiabatic selfconsistent collective coordinate method 
\cite{matsuo00,nakatsukasa12,nakatsukasa16},
Hinohara et al. microscopically
constructed the quadrupole collective Hamiltonian \cite{hinohara10}.
The collective potential is provided by the Hartree--Fock--Bogoliubov
(HFB) calculation with constraints on the $(\beta,\gamma)$ values.
The collective inertial functions are given
as those of normal modes of local quasiparticle-random-phase approximation
(QRPA),
which properly includes time-odd components in the mean field.
The numerical calculation was performed with
the pairing-plus-quadrupole (P+Q) Hamiltonian.
They showed a significant time-odd effect on
collective inertial functions
and on excitation levels in nuclei.
The goal of our work
is an extension of the work by Hinohara et al.
\cite{hinohara10,hinohara11,hinohara12},
that is,
to replace the semi-phenomenological P+Q Hamiltonian by
modern Skyrme EDFs.
It will be also an extension to include $\gamma$ degree of freedom
from the work of 
Yoshida and Hinohara \cite{yoshida11} that constructed a
three-dimensional quadrupole collective Hamiltonian based on
the Skyrme EDF restricted to axially symmetric shapes.

In order to achieve this goal,
the challenge is to perform the local QRPA with a Skyrme EDF
at the constrained HFB states of triaxial shapes.
The axially deformed QRPA calculations with the modern Skyrme, 
Gogny, and covariant EDFs
have recently become available
\cite{yoshida08,peru08,penaarteaga09,losa10,terasaki10}, and
the three-dimensional (3D) RPA calculations without pairing were
achieved \cite{imagawa03,inakura09,inakura11,inakura13}.
However, currently, an efficient framework to solve selfconsistent QRPA
with modern EDFs
applicable to triaxial shapes is still missing,
although there are some related studies using the real-time method
\cite{stetcu11, ebata10, scamps13b, scamps14, ebata14}.
%

In this article, as a first step toward the goal, 
we construct an efficient QRPA solver
for triaxially deformed superfluid nuclei with the Skyrme EDF,
with the help of the finite amplitude method (FAM)
\cite{nakatsukasa07, inakura09, avogadro11, stoitsov11, liang13, niksic13, kortelainen15}.
We start from the HFB code using the two-basis method
of the 3D Cartesian coordinate representation.
We apply the method to multipole modes of excitation in
triaxial nuclei as well as in axially symmetric nuclei for benchmark
and show its feasibility.

\textit{Development of 3D FAM-QRPA.}
%
%
Since the details of the derivation of the FAM equations for QRPA
can be found in Refs. \cite{avogadro11, stoitsov11}, 
we here recapitulate the basic idea and formulae of the FAM.
We start from linear response equation
\begin{subequations}\label{eq:FAM}
\begin{align}
(E_\mu + E_\nu -\omega)X_{\mu\nu}(\omega) + \delta H^{20}_{\mu\nu}(\omega) &= -F^{20}_{\mu\nu} \; ,\\  
(E_\mu + E_\nu +\omega)Y_{\mu\nu}(\omega) + \delta H^{02}_{\mu\nu}(\omega) &= -F^{02}_{\mu\nu}\; ,
\end{align}\end{subequations}
where $X$ and $Y$ are FAM amplitudes at a given frequency $\omega$.
$\delta H^{20(02)}$ and $F^{20(02)}$ are two-quasiparticle matrix elements
of an induced Hamiltonian and an external field, respectively \cite{avogadro11}.

The FAM equation is solved iteratively at each $\omega$.
First, from the $X$ and $Y$ amplitudes at the previous iteration,
the induced density $\delta\rho$ and pairing tensors 
$\delta\kappa$ and $\overline{\delta\kappa}$ are calculated as
\begin{align}
\delta\rho&=UXV^T + V^*Y^TU^\dagger  \; ,\\
\delta\kappa &=U X U^T + V^* Y^T V^\dagger \;, \\ 
\overline{\delta\kappa} &= V^* X^\dagger V^\dagger + U Y^* U^T \;,  
\end{align}
where $U$ and $V$ matrices are taken from the HFB ground state.
The induced pair density has two independent components;
$\delta\kappa$ is proportional to $e^{-i\omega t}$ and the other 
$\overline{\delta\kappa}$ proportional to $e^{i\omega t}$
\cite{avogadro11}.

Next, induced Hartree--Fock (HF) Hamiltonian $\delta h$ and 
pair fields $\delta\Delta$ and $\overline{\delta\Delta} $ are obtained
using a small real parameter $\eta$ as
\begin{subequations}\label{eq:induced-Hamiltonian}
\begin{align}
	\delta h &= \frac{1}{\eta}(h[\rho+\eta\delta\rho] - h[\rho]) 
	 = h^{(1)}[\delta \rho] + \delta h^{(\alpha)}[\rho,\delta\rho]\;, \\
\delta \Delta &= \frac{1}{\eta}(\Delta[\kappa+\eta\delta\kappa] - \Delta[\kappa]) 
         = \Delta[\delta\kappa] \;,\\
\overline{\delta \Delta} &= \frac{1}{\eta}(\Delta[\kappa+\eta\overline{\delta\kappa}] - \Delta[\kappa]) 
         = \Delta[\overline{\delta\kappa}] \;,
\end{align}\end{subequations}
where $\rho$ and $\kappa$ are the density and pair tensor 
in the ground state, respectively.
Most of the terms in the HF Hamiltonian $h$ linearly depends on $\rho$,
while there is a term with density dependence of fractional power
$\rho^\alpha$.
We denote here the former as $h^{(1)}$ and the latter as $h^{(\alpha)}$;
$h=h^{(1)} + h^{(\alpha)}$.
At the last equality for each field in Eq.~(\ref{eq:induced-Hamiltonian}), 
the explicit linearization with respect to induced densities is performed
for $\delta h^{(\alpha)}$, while the rest of the terms can be obtained
by simply replacing $\rho$ by $\delta\rho$.
In this paper, we use the volume-type pairing without density dependence,
thus, $\delta \Delta$ can be calculated as the last equation of
Eq.~(\ref{eq:induced-Hamiltonian}b). 
If the pair field $\Delta$ has a density dependence,
the explicit linearization is required for $\Delta$,
similar to $\delta h^{(\alpha)}$.
In the present scheme \cite{kortelainen15},
the induced fields~(\ref{eq:induced-Hamiltonian}) do not depend on $\eta$
which was required by the original FAM formulation \cite{nakatsukasa07}.

Finally, $\delta H^{20(02)}$ are constructed from $\delta h$,
$\delta\kappa$, and $\overline{\delta\kappa}$, then,
new $X$ and $Y$ amplitudes are obtained from Eq.~(\ref{eq:FAM}).
We employ the modified Broyden method \cite{baran08} for the FAM iterations.
The convergence is reached in about 60--70 iterations at most,
when the convergence condition is set as the maximum difference 
between two successive iterations of $X$ and $Y$ less than
$10^{-7}$;
$|X^{(n)}_{\mu\nu}-X^{(n-1)}_{\mu\nu}|<10^{-7}$ and
$|Y^{(n)}_{\mu\nu}-Y^{(n-1)}_{\mu\nu}|<10^{-7}$ for $\forall\mu\nu$.
The imaginary part of the frequency $\omega$ has been introduced 
as $\omega \to \omega + i\gamma$ with $\gamma=0.5$\,MeV.
The spacing in discretized $\omega$ is taken to be 0.5\,MeV 
to compute strength functions in the following.

The FAM strength function at each $\omega$
is obtained with the converged $X(\omega)$ and $Y(\omega)$ amplitudes as 
\begin{align}
S(\omega) = -\frac{1}{\pi}\text{Im}
\left(\sum_{\mu<\nu} F_{\mu\nu}^{20*} X_{\mu\nu}(\omega)
+ F_{\mu\nu}^{02*}Y_{\mu\nu}(\omega)\right) \;.
\end{align}
We use the one-body external operators as
$\sum_{i=1}^A e_i^{\rm eff} f_{LK}(\boldsymbol{r}_i)$
with $f_{LK}(\boldsymbol{r}_i)=r_i^L Y_{LK}(\hat{\boldsymbol{r}}_i)$ 
and with $f_{00}(\boldsymbol{r}_i) = r_i^2$ for the monopole operator.
The effective charge is adopted as
$e^{\rm eff}=eZ/A$ for the isoscalar operators,
and for isovector operators
$e^{\rm eff}=eZ/A$ ($-eN/A$) for neutrons (protons).
We define the quadrupole operators with
the $x$-signature quantum number of $r_x=\pm 1$ as
$Q_{LK}^{(\pm)}=(f_{2K}\pm f_{2-K})/\sqrt{2}$ for $K>0$.
These operators are written in a simple form in terms of
the Cartesian coordinate $(x,y,z)$ and convenient in the 3D code.
Choosing the $z$ axis as the symmetry axis,
the strength function for $Q^{(+)}_{2K}$ in axially symmetric nuclei 
is identical to that for $Q^{(-)}_{2K}$.
For spherical nuclei,
all the quadrupole operators with different $K,r_x$
carry equal strengths.

We have constructed a 3D FAM-QRPA code based on 
the 3D Skyrme-HFB code \textsc{cr8} \cite{bonche87, gall94, terasaki95},
which is an extension of the 3D HF+BCS code \textsc{ev8} 
\cite{ev8,ev8new}.
The ground state is obtained by the two-basis method \cite{gall94,terasaki95},
where the HF basis that diagonalizes
the HF Hamiltonian and the canonical basis that diagonalizes 
the density matrix are simultaneously used.
The single-particle wave functions are represented on the square mesh
in the 3D Cartesian space 
and eigenstates of 
$z$ signature, parity, and $y$ time simplex.
As a result, each single-particle wave function has
a specific reflection symmetry 
about $x=0$, $y=0$, and $z=0$ planes \cite{bonche85,bonche87,ev8,ev8new,hellemans12}.
We take into account these symmetry properties when
calculating two-quasiparticle matrix elements of the induced densities and
fields in the FAM equations,
which significantly reduce the computational task.
The working volume is then limited to only 1/8 space 
($x>0$, $y>0$, $z>0$)
of the whole volume for both HFB and FAM computations.
The mesh spacing of $\Delta x = \Delta y = \Delta z = 0.8$\,fm is used
in HFB and FAM.
Note that, since the present FAM code can calculate only excitation modes
which conserve parity and $z$-signature symmetries,
the $K=1$ modes of quadrupole excitation ($Q_{21}^{(\pm)}\sim yz$ and $zx$),
which violate the $z$-signature symmetry, cannot be computed.
For this case, we rotate the ground-state wave functions
to switch the labeling of the axes $(x,y,z)$,
so as to make $yz (zx) \rightarrow xy$.
Then, these modes conserve the $z$ signature and the present FAM code
can handle these.

We used SkM$^*$ \cite{bartel82} and SLy4 \cite{chabanat98} parametrizations, 
which have been widely used and known to be stable to QRPA calculations.
We used the volume pairing with a pairing window of 20\,MeV
above and below the Fermi energy in the HF basis
described in Refs. \cite{bonche85,ev8,ev8new}.
We applied the same pairing-cutoff procedure for both HFB and 
FAM-QRPA calculations.
The pairing strength was determined so as to reproduce 
the neutron pairing gap of 1.25\,MeV in $^{120}$Sn.
For simplicity, we used the same pairing strength for 
neutrons and protons.

Before showing the results, we note the treatment of the boundary 
condition in the HFB calculations.
The continuum (positive-energy) HF states in the cubic boundary condition,
which has been used in the codes \textsc{cr8} and \textsc{ev8},
violate the spherical symmetry in our FAM calculation.
For the isoscalar quadrupole modes in the spherical nucleus ${}^{20}$O,
the strengths at energies around the giant resonance
vary depending on $K$ (about $30\%$ difference at most).
To avoid this symmetry-violation effect,
we try to mimic the sphere-type boundary condition,
namely
add an artificial potential to the HF potential,
$V_{\rm sph}(r)=1000$\,MeV at $r>R_{\rm max}$ and 
$V_{\rm sph}(r)=0$ at $r<R_{\rm max}$.
We confirmed that this change in the boundary condition
does not affect the ground-state property.
Furthermore, 
we obtained that the difference in the giant resonance strengths
among different $K$ is at most $4\%$ for ${}^{20}$O,
which is the same order of the deviation observed in the unperturbed strengths
of different $K$.

\begin{figure}[t]
\includegraphics[width=\linewidth]{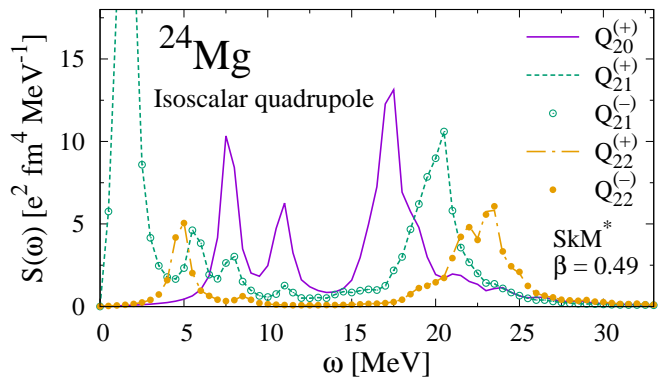}
\caption{\label{fig:strengthMg}
(Color online) Isoscalar quadrupole strengths of different $K$
as a function of $\omega$ for ${}^{24}$Mg calculated with SkM$^*$. 
}
\end{figure}

\textit{Results.}
We first compute isoscalar quadrupole modes of 
an axially symmetric nucleus ${}^{24}$Mg with the SkM$^*$ EDF,
to test our computational code.
We adopt the square mesh space of $15^3$ and
$R_{\textrm{max}}=12.4$\,fm.
The number of HF-basis states is 910
for both protons and neutrons.
We obtained the prolately deformed ground state with $\beta=0.49$.
In this configuration, the pairing vanishes for both neutrons and protons.
Figure~\ref{fig:strengthMg} shows the isoscalar quadrupole strengths 
of ${}^{24}$Mg.
By comparing our result to a previous FAM investigation 
based on the axially symmetric \textsc{hfbtho} in Ref. \cite{kortelainen15},
we found good agreement of the peak energies as well as
the shapes of the strength functions in each $K$.
The widths of the giant resonances for all $K$ in our strengths
are wider than those in Ref. \cite{kortelainen15}.
The peak of $K=1$ spurious mode associated with 
the rotational-symmetry breaking in the ground state
appears at a finite energy 
($\omega\approx 1.5$\,MeV).
This deviation from zero energy
is due to the use of the finite mesh size, 
which was extensively discussed in Ref. \cite{terasaki10}.
The energy-weighted sum-rule (EWSR) values summed up to $\omega=50$\,MeV
are exhausted by $98.5\%$ ($K=0$) and $98.3\%$ ($K=2$).
The strengths with $r_x=\pm 1$ coincide for $K=1$ and 2.

\begin{figure}[t]
\includegraphics[width=\linewidth]{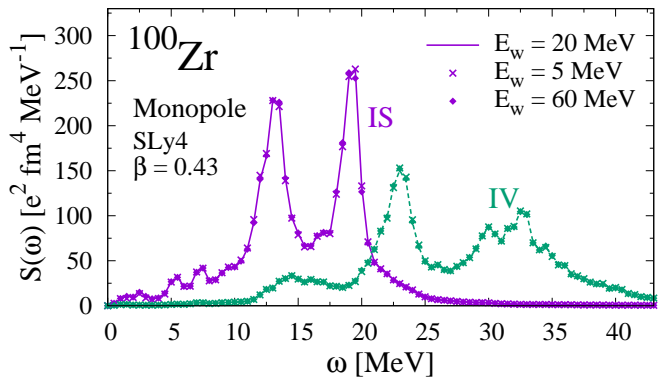}
\caption{\label{fig:strengthZr}
(Color online) Isoscalar (IS) and isovector (IV) monopole strengths 
as a function of $\omega$ for ${}^{100}$Zr with SLy4
and with different pairing window energies $E_w=20$\,MeV (solid line),
5\,MeV (cross), and 60\,MeV (diamond).
}
\end{figure}

Figure~\ref{fig:strengthZr} shows isoscalar and isovector monopole strengths 
in a prolately deformed superfluid nucleus ${}^{100}$Zr
computed with $17^3$ mesh ($R_{\textrm{max}}=14.0$\,fm) 
and 1120 HF-basis states with SLy4 EDF.
The obtained ground state has finite pairing gap for protons
(normal phase in neutrons) and $\beta=0.43$.
Compared with previous axial matrix-form QRPA \cite{yoshida10}
and FAM-QRPA \cite{stoitsov11},
nice agreement on the peak energies is obtained,
even though we used different pairing functionals and different pairing cutoff
from those in Refs. \cite{yoshida10, stoitsov11}.
In Fig.~\ref{fig:strengthZr},
we also show the dependence on the pairing window energy.
The pairing strength of each pairing window was adjusted
with the method mentioned above.
No significant dependence of pairing window energy is observed in the strengths.
Furthermore, the $0^+$ spurious modes corresponding to the pair rotation
are not seen in the monopole strength function.
This indicates good decoupling between the pair and monopole modes of
excitation.

\begin{figure}[t]
\includegraphics[width=\linewidth]{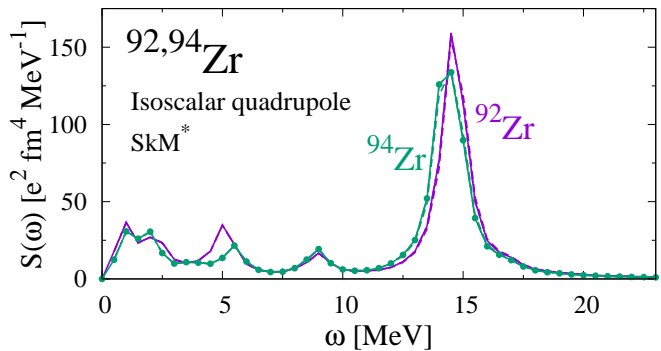}
\caption{\label{fig:strengthZr9294}
(Color online) Isoscalar quadrupole strengths for spherical nuclei ${}^{92}$Zr (purple)
and ${}^{94}$Zr (green).
All the strengths with $K=(0,1,2)$ (solid, dashed, and dot-dashed lines) 
and $r_x=\pm 1$ are identical.
}
\end{figure}

We show the isoscalar quadrupole strengths of ${}^{92}$Zr and ${}^{94}$Zr
in Fig.~\ref{fig:strengthZr9294},
which are next to an LLFP ${}^{93}$Zr.
The model space was same as in ${}^{100}$Zr, but SkM$^*$ EDF was used.
The ground states of ${}^{92}$Zr and ${}^{94}$Zr are spherical and superfluid 
in both neutrons and protons.
Since these nuclei are spherical in their ground state,
the strengths of different $K$ agree with each other.
The giant resonance peaks appear at around 15\,MeV, while
we also observe that the lowest peak is located at about $\omega\approx 1$ MeV.
These low-energy modes are expected to play an important role
in the shape fluctuation, which will be our future target.

\begin{figure}[t]
\includegraphics[width=\linewidth]{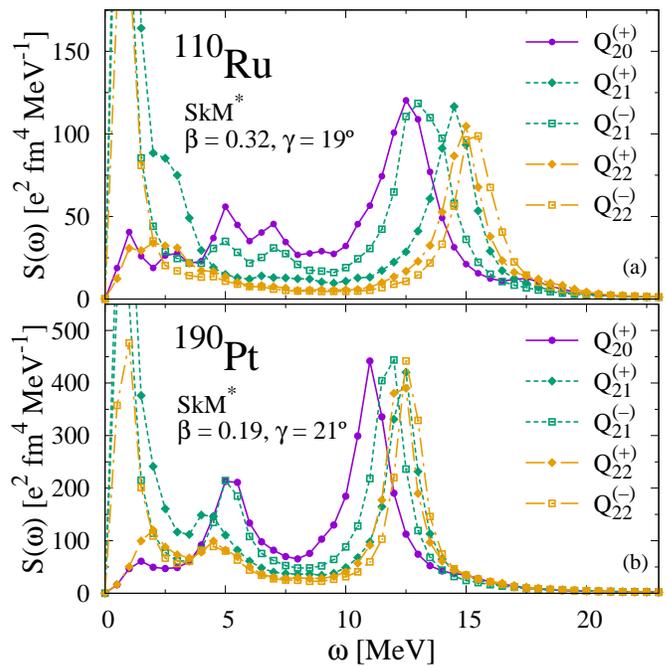}
\caption{\label{fig:strengthRu-Pt}
(Color online) Isoscalar quadrupole strengths for 
triaxial nuclei ${}^{110}$Ru (a) and ${}^{190}$Pt (b).
}
\end{figure}

Finally,
we show  
the isoscalar quadrupole modes in triaxially deformed superfluid nuclei.
A typical mass region of appearance of triaxial ground state
is the $A\approx 100$ region and Pt isotopes \cite{scamps14, ebata14}.
We take here ${}^{110}$Ru and ${}^{190}$Pt, calculated with the SkM$^*$ EDF. 
We set the longest, middle, and shortest axes to be
$z$, $x$, and $y$ axes, respectively.
Note that the magnetic quantum number $K$ is not good quantum number 
for triaxial nuclei. 
For our convenience, however, we use the $K$ values
to specify the type of quadrupole operators $Q_{2K}^{(\pm)}$.

Figures~\ref{fig:strengthRu-Pt}(a) and \ref{fig:strengthRu-Pt}(b) show 
the isoscalar quadrupole strengths
of ${}^{110}$Ru and of ${}^{190}$Pt, respectively.
In both nuclei, neutrons are in the superfluid phase, while protons are not.
The obtained quadrupole deformation parameters are 
$\beta=0.32$, $\gamma=19^\circ$
for ${}^{110}$Ru with the same model space as in Fig.~\ref{fig:strengthZr9294}.
For ${}^{190}$Pt with an enlarged model space as $19^3$ mesh 
($R_{\textrm{max}}=15.6$\,fm) and 1360 HF-basis states,
we obtain the ground state with $\beta=0.19$, $\gamma=21^\circ$.
In both nuclei,
our calculation clearly produces additional signature splitting
of the strength
in which the peaks with different $x$ signature no longer coincide,
due to the triaxial deformation. 
We obtain three spurious modes near zero energy
due to the existence of rotations around $x$, $y$, and $z$ axes.
The EWSR values are well satisfied, 98--99\% for $K=0$ and $(K,r_x)=(2,+)$ modes.
We have also confirmed that the isoscalar quadrupole response has
no significant dependence on the pairing window
in a range from 10 to 50\,MeV.


\begin{table}[t]
\caption{\label{tab:convergence}%
EWSR values 
summed up to $\omega=50$\,MeV
for the isoscalar quadrupole $K=0$
strength in units of MeV\,$e^2$\,fm$^4$
in ${}^{110}$Ru with different numbers of HF-basis states $N_{\textrm{HF}}$
and of two-quasiparticle states $N_{\textrm{2qp}}$.
The last two rows show the maximum quasiparticle energies $E_{\textrm{QP}}^{\textrm{max}}$ (in MeV) 
of neutrons (n) and protons (p). 
}
\begin{ruledtabular}
\begin{tabular}{rrccc}
$N_{\textrm{HF}}$ & $N_{\textrm{2qp}}$ & EWSR & $E_{\textrm{QP,n}}^{\textrm{max}}$ &$E_{\textrm{QP,p}}^{\textrm{max}}$\\
\hline
240  & 7400  & 6913 &  41.3 & 34.2 \\
440  & 24650 & 7121 &  41.3 & 39.8 \\
728  & 67130 & 7159 &  41.3 & 49.5 \\
910  & 104713& 7162 &  45.9 & 55.3 \\
1120 & 158368& 7164 &  50.6 & 60.2
\end{tabular}
\end{ruledtabular}
\end{table}

For the isoscalar quadrupole $Q_{20}$ of ${}^{110}$Ru,
we examine in detail the convergence property of EWSR
with respect to the number of HF-basis states.
Table~\ref{tab:convergence} shows 
the calculated EWSR values with FAM.
For neutrons in $N_{\textrm{HF}}\leq 728$ and protons in $N_{\textrm{HF}} = 240$,
the main component of the highest quasiparticle state is the deepest hole state
in the HF basis.
We reach an approximate convergence of EWSR at $N_{\textrm{HF}}\geq 728$
which corresponds to the number of two-quasiparticle states
$N_\textrm{2qp}\geq 67,130$.
In the present 3D calculation, even though the size of the space is reduced
by incorporating the parity and the $z$-signature symmetry,
the number of two-quasiparticle states easily exceed 100,000.
It requires enormous computational power and memory capacity to
explicitly construct such large QRPA matrices.
The FAM significantly reduces the computational burden and
provides a feasible numerical approach to the QRPA.

\textit{Conclusions.}
Toward fully microscopic and non-empirical construction of
the five-dimensional quadrupole collective Hamiltonian,
we have developed a 3D FAM-QRPA code
applicable to triaxially deformed nuclei with superfluidity. 
We demonstrated that the results showed good agreement 
with the previous axial QRPA results on multipole modes of excitation
in axially symmetric nuclei,
${}^{24}$Mg and ${}^{100}$Zr.
In axially deformed nuclei, the quadrupole strength functions
with the same $K$ but different $x$ signature $r_x=\pm 1$ 
coincide to each other.
The rotational zero-energy modes around $x$ and $y$ axes exist,
but that around the $z$ (symmetry) axis does not.

We applied our 3D FAM-QRPA to isoscalar quadrupole modes 
in triaxially deformed superfluid nuclei, ${}^{110}$Ru and ${}^{190}$Pt.
Five different peaks in the strength functions appear
depending on $K$ and the signature $r_x$.
Three rotational modes also emerge at zero energy, associated with
rotations about all the three axes $(x,y,z)$
because of the triaxial deformation.

The present FAM computation depends mainly on 
the numbers of the mesh points and of HF-basis states.
The computation of the isoscalar quadrupole strength for
100\,$\omega$ points in Fig.~\ref{fig:strengthRu-Pt}(b)
is about 200\,CPU hours in total and 3.5\,GB memory.
This indicates the efficiency of our computational method
and feasibility in currently available computational resources.

We intend to develop a parallelized local QRPA computer code
based on the present
FAM-QRPA framework,
to derive the collective inertial functions 
at every $(\beta, \gamma)$ point.
%
To obtain low-lying discrete normal modes in the local QRPA,
the contour integration technique of Ref. \cite{hinohara13} may be useful.
The extension of the present FAM-QRPA to the local QRPA 
is in progress.

\textit{Acknowledgments.}
The authors acknowledge Nobuo Hinohara for fruitful discussions.
This work was funded by ImPACT Program of
Council for Science, Technology and Innovation
(Cabinet Office, Government of Japan).
Numerical calculations were performed in part using the COMA (PACS-IX) 
at the Center for Computational Sciences, University of Tsukuba.



%

\end{document}